\documentclass[sigconf,nonacm]{acmart}

\AtBeginDocument{%
  }

\newcommand{\tagtype}[1]{{\tt{#1}}}
\newcommand{\codevar}[1]{{\tt{#1}}}
\newcommand{\tagrule}[1]{{\tt{#1}}}

\begin{document}

\title[Pipe-Cleaner]{Pipe-Cleaner: Flexible Fuzzing Using Security Policies}

\author{Allison Naaktgeboren}
\email{naak@pdx.edu}
\orcid{0009-0004-0405-9306}
\affiliation{
  \institution{Portland State University}
  \city{Portland}
  \state{OR}
  \country{USA}
}

\author{Sean Noble Anderson}
\email{ander28@pdx.edu}
\affiliation{%
  \institution{Portland State University}
  \city{Portland}
  \state{OR}
  \country{USA}
}

\author{Andrew Tolmach}
\email{tolmach@pdx.edu}
\affiliation{%
  \institution{Portland State University}
  \city{Portland}
  \state{OR}
  \country{USA}
}

\author{Greg Sullivan}
\email{gsullivan@draper.com}
\affiliation{%
  \institution{Charles Stark Draper Laboratory}
  \city{Cambridge}
  \state{MA}
  \country{USA}
}

\renewcommand{\shortauthors}{Naaktgeboren et al.}

\begin{abstract}
  Fuzzing has proven to be very effective for discovering certain classes of software flaws, but less effective in helping developers process these discoveries. Conventional crash-based fuzzers lack enough information about failures to determine their root causes, or to differentiate between new or known crashes, forcing developers to manually process long, repetitious lists of crash reports.
  
Also, conventional fuzzers typically cannot be configured to detect the variety of bugs developers care about, many of which are not easily converted into crashes. 

To address these limitations, we propose Pipe-Cleaner, a system for detecting and analyzing C code vulnerabilities using a
refined fuzzing approach.
Pipe-Cleaner is based on flexible developer-designed security policies enforced by a tag-based runtime reference monitor,
which communicates with a policy-aware fuzzer.
Developers are able to customize the types of faults the fuzzer detects and the level of detail in fault reports.
Adding more detail helps the fuzzer to differentiate new bugs, discard duplicate bugs, and improve the clarity of results for bug triage. We demonstrate the potential of this approach on several heap-related security vulnerabilities, including classic memory safety violations and two novel non-crashing classes outside the reach of conventional fuzzers: leftover secret disclosure, and heap address leaks. 
\end{abstract}



\settopmatter{printfolios=true}

\maketitle

\section{Introduction}\label{sec:intro}

Fuzzing\cite{30yrfuzzing-Miller2022,OSSfuzz-5m-trophies,afl-trophycase,libfuzzer-trophycase,rustfuzz-trophycase,honggfuzz-trophycase}, also known as fuzz testing, is a dynamic, probabilistic software-testing technique. It attempts to thoroughly explore the input space of the target program, searching for inputs that cause ``interesting behavior,'' which, in most production fuzzers, is hard-coded to mean crashes (segmentation faults) and hangs.
Although this simple definition has historically worked well for bug discovery, it limits the fuzzer's ability to detect duplicate faults, aid in bug report triage, and detect non-crash bugs. We identify three key problems with standard fuzzing approaches.

\emph{The (De)Duplication Problem.}
Fuzzing results have a bad signal to noise ratio. Due to the limited information in a crash report, the default mechanism for differentiating crashes is to compare hashes of their stack traces, which is fast but can cause the fuzzer to both over-report and under-report crashes \cite{Klees18:Evaluating}. They over-report when crashes with the same root cause have different hashes (which is very common), and under-report when two unrelated crashes accidentally have matching hashes.

Over-reporting can lead to many duplicates clogging the results given to developers, reducing time spent on bugfixes~\cite{IGOR-crashdedup-Juang21}. 

\emph{The Crash Triage Problem.}
Basic bug triage requires three things: (1) the cause of failure, (2) whether there are security implications, and (3) which developer teams are responsible for the fix (often approximated by locations in the source code). Current fuzzers cannot provide most of this information, so their reports cannot be easily triaged, and hence are liable to be ignored. More than half of the fuzzer crashes reported to the Linux kernel are ignored \cite{7yrsLinuxKernelFuzz23}, likely due to lack of information in the crash report \cite{linuxtriagetroubles22}. This is one of the top concerns of fuzzer users \cite{Nourry-humanfuzzingchals23}.

\emph{The Crash Bias Problem.}
The reliance on crashes also biases fuzzing results towards flaws that can be easily signaled by crashes, such as memory corruption, leaving other types of security flaws undetected. This is part of a wider problem of inflexibility in fuzzer design.

There are typically two ways to tailor a fuzzer to new classes of bugs: either write a new fuzzer, or add a sanitizer to the executable being fuzzed. Most sanitizers work by inserting conditional crashing code during a compiler pass, a difficult and demanding task \cite{google-fuzz-beyond-membugs}. Sanitizers are typically incompatible with each other \cite{sanitizers-Peko21}, difficult to modify, and are unavailable under certain conditions.
While some sanitizers, such as ThreadSanitizer \cite{clangThreadSanitizer}, do produce helpful log output, fuzzers are unaware of this, and neither capture nor use any extra information the sanitizer might emit. The typical workflow for fuzzing with sanitizers is to compile with a sanitizer, fuzz, and then run crashing inputs individually on a sanitized binary and hope that a useful error message results.

These problems all fundamentally stem from relying on crashes and crash dumps as the sole means of discovering
and reporting faults. To address them, we propose \emph{Pipe-Cleaner}, a new approach to fuzzing C code
that executes the target program under control of a security \emph{reference monitor}~\cite{Anderson72:PlanningStudy}.
Specifically, we use the Tagged C system~\cite{TaggedC-RV23}, which enforces
arbitrary user-configurable security policies based on metadata tags carried for
each value. 
Security policies can range from classic static and dynamic memory safety~\cite{eternal13} to
fine-grained information flow control~\cite{Denning76:SFIlattice} supporting data confidentiality or integrity properties. 
Any policy violation causes the fuzzer to be notified with a report that includes dynamic context information
which can be used to help de-duplicate and classify bugs. 

Thus, developers can focus fuzzing resources on the bugs they care about and receive nuanced information about faults for better triage. Operators can swap out policies without needing a new fuzzer (or a new compiler pass), or choose to fuzz with multiple policies at once.

Tagged C can be thought of as a highly configurable sanitizer, and like other fuzzing approaches based on 
code instrumentation, Pipe-Cleaner deliberately trades off execution speed against improved quality of fault information.
Currently, Tagged C is available only as an interpreter, but a faster execution engine based on source-to-source
insertion of instrumentation is under development, and ultimately we hope 
to use the hardware tagging support known as Processor Interlocks for Policy Enforcement (PIPE)~\cite{Dhawan+15,Azevedo+16,Azevedo+15}\footnote{Variants of PIPE have also been called PUMP~\cite{Dhawan14:PUMP}, SDMP~\cite{Dover16}, or CoreGuard~\cite{Dover20}.}, 
which monitors protected metadata tags in parallel with ordinary execution on values to obtain performance comparable to
normal code. For this reason, Pipe-Cleaner policies operate on metadata tags rather than on actual values.  This is one of
the features that distinguishes our approach from property-based testing (PBT)~\cite{Lampropoulos19:CoverageGuidedPBT},
which relies on inspecting values. Another difference is that PBT is usually employed to check program-specific functional
properties rather than generic security properties. 

In summary, we make the following contributions:
\begin{enumerate}
    \item We present Pipe-Cleaner, a novel fuzzing system using flexible developer-designed security policies (Section~\ref{sec:design}).
    \item We show examples of how Tagged C policies enrich fuzzer behavior through improved duplicate detection and more precise bug reports (Section~\ref{sec:policies}).
    \item We implement and evaluate these example policies against intuitive metrics
           for characterizing fuzzer behavior on nontraditonal bug classes (Section~\ref{sec:m-and-m}).
\end{enumerate}

An artifact including everything necessary to reproduce the behavior of our prototype implementations will be made available
as part of the final version of this paper.

\section{Design of the Pipe-Cleaner System}\label{sec:design}

\begin{figure}[tb]
  \centering
  \includegraphics[width=\linewidth]{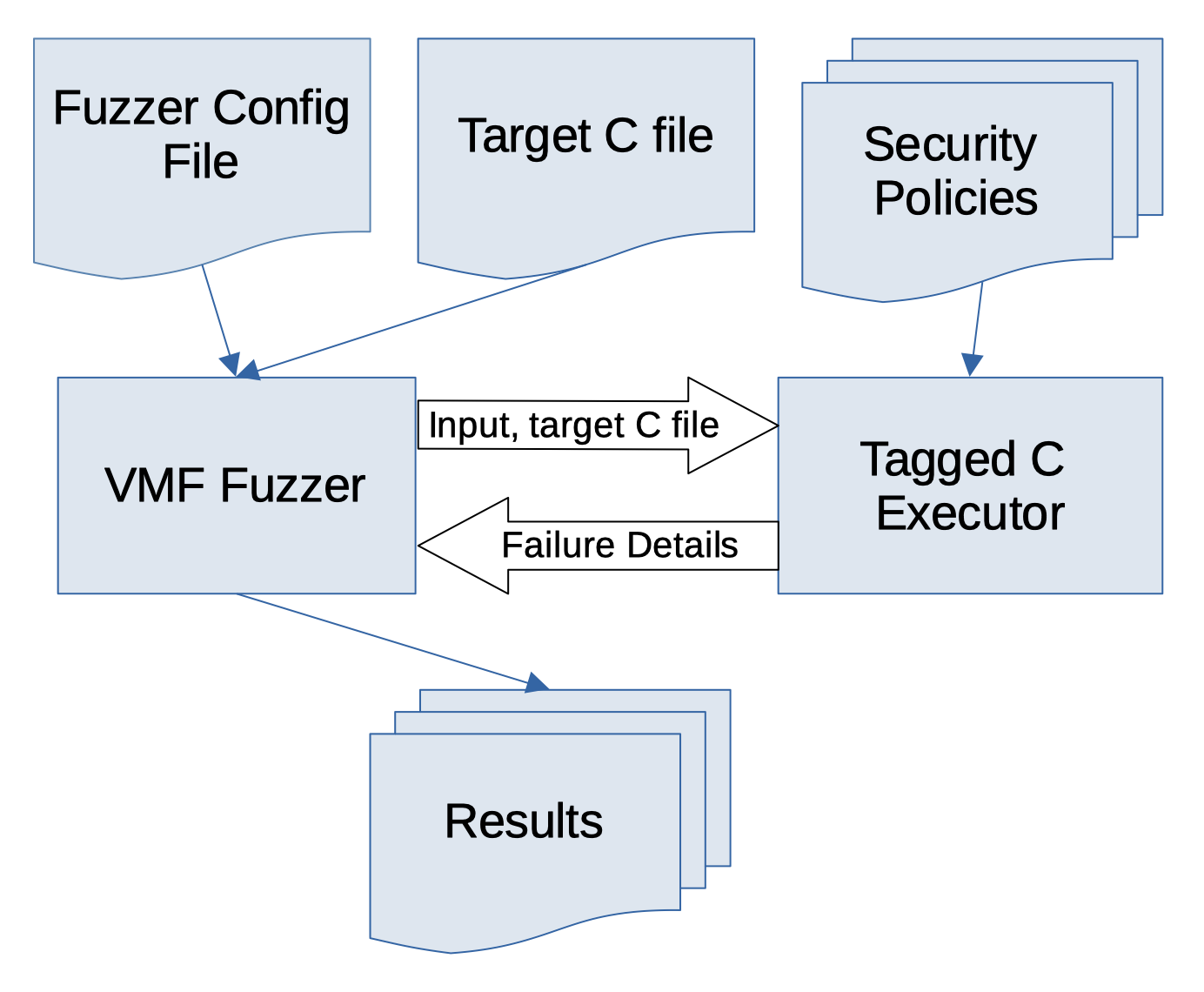}
  \caption{The Pipe-Cleaner System}
  \Description{flow chart of Pipe-Cleaner System}
  \label{fig:pipecleanerdiagram}
\end{figure}

Pipe-Cleaner (Figure~\ref{fig:pipecleanerdiagram}) takes a target C program, a configuration file, and a user-defined runtime security Tagged C {\em policy}---a set of rules restricting the behavior of the program~\cite{TaggedC-RV23}.  It runs the target in an execution engine (an interpreter, in our current prototype) that enforces the policy. If an execution would violate the policy, it instead halts (referred to as a {\em failstop}). These executions are run inside of a modified off-the-shelf fuzzer which consumes the failstops much like any other crash, but extracts significantly more data for use in triage.

Our fuzzing harness is a modification of VMF \cite{VMF}.

\paragraph{Tagged C}

Tagged C is a C variant in which every value is paired with a piece of metadata termed a {\em tag}, representing information of import to the policy: ``this value belongs to Bob,'' or ``this value is a pointer to object \codevar{x},'' for example. An additional piece of metadata is associated with the current program control state. At key points in execution, termed {\em control points}, Tagged C checks the tags on relevant values and the state against a {\em tag rule}. The rule either determines updated tags or causes execution to failstop. A Tagged C policy can encode many different kinds of security policies, as long as they can be expressed in terms of the flow of tags through the program state. It is limited in its ability to directly access values, but dealing only with metadata enables a range of important policies, including memory safety and various information-flow policies. Tagged C is currently implemented as a reference interpreter that models the tag system in software, with some limitations discussed in Section \ref{sec:implimits}. 

\paragraph{Policies}
Policies are the foundation of Pipe-Cleaner. They work with the fuzzer by identifying security violations and providing the detailed diagnostic information used to make a unique key for the violation.

Policy designers may choose to enforce several security guarantees in a single policy, as shown in the HeapSafety policy (Section \ref{sec:policies}), or focus narrowly on one, as shown in the DoubleFree policy (Section \ref{sec:doublefree}). We expect that (appropriately configured) Pipe-Cleaner policies will detect classic fuzzing bugs, such as heap overread or overwrite, as well as or better than a fuzzer that only detects bugs via segfaults, and enable us to find novel classes of bugs as well.

In the current system, policies are written in the programming language Gallina, part of the Coq theorem prover \cite{coqproofasst}.

Each policy defines sets of possible tags, functions for each of the control points specified in the Tagged C API~\cite{TaggedC-RV23}, and the initial tag state of the system.

There are three types of tags available to policy designers: value, location, and control. A \emph{value tag}
is associated with each value flowing through the program via variable reads and writes and expression evaluation;
for example, a policy might use two value tags to distinguish pointer values from integer values.
A \emph{location tag} is associated with each address in the program that contains accessible data, i.e. each slot
in an array or structure, and each address-taken variable; for example, a location tag might identify the array to
which a particular address belongs.
A \emph{control tag} is associated with the current control state, which changes as execution proceeds from one program point to another; for example, a control tag might track the name of the currently executing function.

The interpreter can run multiple policies at once, using the Cartesian product of tags so that policies do not interfere with each other. If any policy failstops, so does the combined policy. This facility supports a policy designer who might desire to fuzz with several small tailored policies rather than a large single policy.

\paragraph{VMF}
The modified VMF fuzzer forms the final part of Pipe-Cleaner. There are new modules for initialization, execution, processing feedback, and results. The default storage, controller, input generator, output storage, and mutation modules are used. For the purposes of this paper, the feedback module, PipeCliInterpreterFeedback, is the most important. It determines if duplication has occurred and processes the results of a fuzz run. It processes the detailed error and optional log file from the Tagged C executor to determine if the bug is new or known. The module uses the detailed feedback to form an identity for the bug, based on its type, which policy reported it, and which source locations were involved. Some bug classes might not require details to identify; others, like double free, require multiple details to uniquely identify for deduplication and triage. If the bug is new, the module minimally processes it, adds its identity to its map of known bugs, and saves the input. If it is a known bug, already present in the map, it updates that bug's duplicate counter and discards the input. For certain bugs, such as those in  Section~\ref{sec:dumpsterdiving}, the feedback module determines whether or not the bug was dangerous.

\section{Tagged C Policies for Heap Fuzzing}
\label{sec:policies}
In this section, we describe the heap policies used in our prototype evaluation.  We focus on the heap because memory corruption makes up the overwhelming majority of vulnerabilities \cite{Lord-memsafety-usgov-23}, and the majority of memory corruption vulnerabilities are heap related \cite{70percentbugarememsafety-chrome}. However, our approach can extend to other aspects of
program security. In each case, we discuss the motivation behind the policy, the tags chosen and the control points of interest.

We introduce the features of Tagged C gradually, as they become needed for the different policies. 
To improve readability, we present policies using pseudo-code rather than actual Gallina.

\subsection{Detecting Double Frees} 
\label{sec:doublefree}

    A double free is a temporal safety violation that occurs when memory in the heap is incorrectly returned to the system, or freed, twice. It is a specialized case of the Use-After-Free (UAF) vulnerability class, often occurring in complicated clean-up routines \cite{OWASP-dfree-def}. Double frees can lead to arbitrary code execution, including the more dangerous arbitrary remote code execution \cite{dropbear-dfree-cve,openssh-dfree-poc}. Double frees are considered highly exploitable by the MITRE Common Weakness Enumeration (CWE) \cite{CWE-dfree-def}. 

Detecting the conditions for a double free vulnerability is fairly straightforward for a tagged system. Thus they make a good introduction to Tagged C security policies.

The DoubleFree security policy focuses on tracking the state of freed memory in the heap.
It does not detect all possible heap corruptions, but instead focuses on finding and reporting double frees. Effective remediation and deduplication requires both \codevar{free()} locations. This is reflected in the tag choices, listed at the top of Figure~\ref{fig:DoubleFreePseudocode}. DoubleFree uses only location tags, and ignores value or control tags.
The memory manager is assumed to be a typical one in which each object has a header that
contains its size (and perhaps other metadata). We use the location tag on the header to represent the status of the overall object:
either \tagtype{AllocatedHeader} for a currently active object, or \tagtype{FreedHeader(srcpos)} for an object that was most
recently freed at source position \codevar{srcpos}. All other memory locations are labeled \tagtype{NotHeader}; at program start-up,
every location has this tag.
The policy interacts with the memory manager at control points corresponding to \codevar{malloc()} and \codevar{free()} calls.
Figure~\ref{fig:DoubleFreePseudocode} shows the rules executed for this policy at these control points.
Each rule takes as inputs zero or more tags on relevant values and pointers, and either assigns as outputs (written as primed variables) a
relevant set of tags on results or \codevar{raise}s a failstop condition.
In reality, these rules take and return additional tags; for simplicity of presentation, we omit these when they are ignored or passed on unchanged by the policy.
Here, \tagrule{MallocT} assigns \tagtype{AllocatedHeader} as the location tag \codevar{hdr'} that will be attached to the header of the newly allocated object.

\tagrule{FreeT} takes the source position \codevar{srcpos} of the \codevar{free()} call and the header location tag \codevar{hdr}, and inspects the latter. 
If this indicates an active object (\tagtype{AllocatedHeader}), the rule updates the tag to \tagtype{FreedHeader(srcpos)}, thus remembering
the source position of this \codevar{free()}. 
If the tag is \tagtype{freedHeader(srcpos)}, the rule detects that this is a double free, and reports
the source position of the first free as well as the current one.
Finally, if the tag indicates that this is not a header at all (\tagtype{NotHeader}), the rule failstops with an error message.
(Although detecting nonsense frees is not a specific goal of this policy, it ``comes for free.'')

\begin{figure}[tb]
  \centering
  \small
\begin{verbatim}
ltags := {AllocatedHeader, FreedHeader(srcpos), NotHeader}
vtags := {}, ctrltags := {}

Function MallocT() :=
 hdr' := AllocatedHeader 

Function FreeT (srcpos, hdr) :=
 case hdr of
  AllocatedHeader =>
   hdr' := FreedHeader(srcpos)
  FreedHeader(ff) =>
   raise Fail("Double free: 1st free {ff}, 2nd free {srcpos}")
  NotHeader => 
   raise Fail("Nonsense free or corrupted pointer at {srcpos}")
\end{verbatim}
  \caption{Key Rules in the DoubleFree Policy}
  \Description{Pseudocode of \tagrule{MallocT} and \tagrule{FreeT} rules for DoubleFree Policy}
  \label{fig:DoubleFreePseudocode}
\end{figure}

\begin{figure}[tb]
  \centering
  \small
\begin{verbatim}
---------------------------------
Unique Sec Policy Failures: 3
Total (nonunique) Failures: 2633
Unique Standard Crashes   : 0
Total (nonunique) Crashes : 0
Total Testcases Executed  : 5765
---------------------------------
Problem Root Cause:
  Policy Violated: DoubleFree.
  Failed Rule: FreeT detects two frees.
  Memory first freed at location …/file.c:81 
  was freed again at location …/file.c:83
TC Filename/ID : 3
Testcase/Input : 22p?

Problem Root Cause:
  Policy Violated: DoubleFree.
  Failed Rule: FreeT detects two frees.
  Memory first freed at location …/file:69 
  was freed again at location …/file.c:75
TC Filename/ID : 5
Testcase/Input : 2?0?

Problem Root Cause:
  Policy Violated: DoubleFree.
  Failed Rule: FreeT detects two frees.
  Memory first freed at location …/file.c:67 
  was freed again at location …/file.c:75
TC Filename/ID : 21
Testcase/Input : "?H]U
\end{verbatim}
\caption{Sample Pipe-Cleaner Output for DoubleFree policy}
\Description{Text output of three different double frees and summary of execution run}  
\label{fig:DoubleFreeOutputExample}
\end{figure}

To secure the heap we must understand the behavior of the memory manager. The default configurations of most modern memory managers are concerned with performance, not security. They typically manage freed memory via a linked data structure, use metadata headers to track the size of the allocation, pad allocations to alignment boundaries, and do not clean or zero memory at \codevar{malloc()} or \codevar{free()}. Several vulnerability classes are built around these assumptions, so a fuzzing policy hoping to catch them must take account of them as well. Security-conscious memory manager features are available, but the typical default configurations on Linux, OSX, Android \cite{glibc-malloc-impl-18}, and Windows \cite{msft-malloc-impl} do not use them.

We choose not to include stack protection in the HeapSafety policy. Other work has demonstrated how to protect the stack in a tagged architecture \cite{Roessler18:Stack,Anderson23:StackSafety}, and we expect the Tagged C version of those policies to be straightforward.

HeapSafety's tags and (simplified) allocation-related rules are shown in Figure~\ref{fig:HeapSafetyPseudocode}. 
This policy uses location, value, and control tags.
The key idea is to identify each allocated heap object with a unique integer identifier, called a \emph{color}~\cite{Clause07:MemsafeTainting,micropolicies-tagmonitors-Amorim15}. Each location associated with the object is tagged with the color. We further distinguish the header location (\tagtype{AllocatedHeader}), the data bytes within the object (\tagtype{Allocated}
for initialized data or \tagtype{AllocatedDirty} for uninitialized data), and any padding bytes (\tagtype{AllocatedPadding}); this supports better error messages. The flexibility of Tagged C's policies 
helps us neatly avoid potential issues with undetected small overflows of sub-objects that can happen in other systems (such as CHERI~\cite{Woodruff19-Cheri-concentrate}). Other locations are marked as being inside or outside the heap.
At the start of the program, all heap locations are tagged with \tagtype{UnallocatedHeap} and the rest of memory with \tagtype{NotHeap}.
Value tags are used to distinguish heap pointers, marked with the color of the object to which they point, from all other values (including pointers outside the heap).
The control tag is used to keep track of the next available color. 

The rules for \codevar{malloc()} and \codevar{free()} control points are more complex for this policy than for DoubleFree.
\tagrule{MallocT} consumes the current control tag (\codevar{pct}) to obtain the next free color \codevar{c}, and
increments it.  It also sets many other tags: the value tag of the the returned pointer (\codevar{pt'}) is marked
as a pointer with color \codevar{c},  the location tags on the freshly allocated region's header, data bytes, and padding
(\codevar{hdr'},\codevar{lts'},\codevar{pad'})
are set appropriately (with the data bytes being tagged \tagtype{AllocatedDirty} because the data has not been initialized yet), and the value tag (\codevar{vt'}) for the data bytes is set to \codevar{NotHeapPointer}.
\tagrule{FreeT} takes the tag on the pointer being freed (\codevar{pt}) as well as the location tag on the header of the object pointed to (\codevar{hdr}), and checks that their colors match; if so, the header location tag is reset to \codevar{UnallocatedHeap}.
There is also an additional rule \tagrule{ClearT} that is executed on
\codevar{free()} operations for each byte in the freed region, and resets its location tag to \tagtype{UnallocatedHeap}
after checking for possible corruption. 

HeapSafety also uses additional rules \tagrule{LoadT} (Figure~\ref{fig:HeapSafetyPseudocode2}) and \tagrule{StoreT} (not shown) to monitor reads and writes to memory. \tagrule{LoadT} is passed the value tag \codevar{pt} of the pointer being loaded from, the address \codevar{addr}
being loaded from, and the list of location tags \codevar{lts} of the bytes being read.
A load succeeds if the memory block tags are allocated with the matching color.
If \tagrule{LoadT} sees \tagtype{AllocatedDirty}, it will log the event but continue execution and the fuzzer will later decide if there was a vulnerability or just a bug, for reasons discussed in Section \ref{sec:dumpsterdiving}.
If the tags are anything else, the operation is a heap overread (or the pointer has been corrupted to lie outside the heap entirely) and a failstop occurs.
\tagrule{StoreT} is very similiar, except that \tagtype{AllocatedDirty} bytes are converted to \tagtype{Allocated} when they are overwritten. 
Loading and storing through \tagtype{NotHeapPointer} values that point into the heap indicates corruption and a failstop occurs.

Finally, there are control points and rules (not shown) corresponding to arithmetic operations and casts on values; these 
propagate the \tagtype{HeapPtr} tags through operations that make sense on pointers (e.g. addition with a constant)
and otherwise set the result value tag to \tagtype{NotHeapPointer}.  Note that this scheme permits a pointer to keep its
color even if it no longer points within the corresponding object; this is acceptable, because any attempt to actually
access memory through the pointer will fail.

\begin{figure}[tb]
  \centering
\small
\begin{verbatim}
ltags := {AllocatedHeader(srcpos,c), Allocated(srcpos,c),
          AllocatedDirty(srcpos,c),
          AllocatedPadding(srcpos,c),
          UnallocatedHeap, NotHeap}
vtags := {HeapPtr(srcpos,c), NotHeapPointer}
ctrltags := {NextId(c)}

Function MallocT(srcpos, pct) :=
 pt' := HeapPtr(srcpos,c)
 hdr' := AllocatedHeader(srcpos,c)
 lts' := AllocatedDirty(srcpos,c)
 pad' := AllocatedPadding(srcpos,c)
 vt' := NotHeapPointer
 pct' := NextId(c+1)
 where pct = NextId(c)

Function FreeT (srcpos, pt, hdr) :=
 case pt, hdr of
  HeapPtr(_,pc), AllocatedHeader(_,hc) => 
   if pc != hc then raise 
     Fail("Corrupted:Free ownership mismatch @{srcpos}")
   else hdr':= UnallocatedHeap 
  HeapPointer(_,_), _ => raise 
   Fail("Nonsense free(corrupted pointer) @{srcpos}")
  _, _ => raise 
   Fail("Attempt to free non-pointer @{srcpos}")

Function ClearT (srcpos, pt, lt) :=
 case pt, lt of
  HeapPointer(_,pc), Allocated(_,oc) |
  HeapPointer(_,pc), AllocatedDirty(_,oc) |
  HeapPointer(_,pc), AllocatedPadding(_,oc) => 
   if (pc != oc) then raise 
    Fail("Corrupted:Clear ownership mismatch @{srcpos}")
   else lt' := UnallocatedHeap
  _, _ => raise 
    Fail("Corrupted: Corrupted data @{srcpos}")
\end{verbatim}
\caption{Allocation-related Rules in the HeapSafety Policy}
\label{fig:HeapSafetyPseudocode}
\Description{Pseudocode of \tagrule{MallocT}, \tagrule{FreeT}, \tagrule{ClearT} rules for HeapSafety}
\end{figure}

\begin{figure}[tb]
  \centering
  \small
  \begin{verbatim}
Function LoadT (srcpos, pt, addr, lts) :=
 case pt of
  HeapPointer(_, pc) => 
   for each lt in lts 
    case lt of 
      NotHeap |
      UnallocatedHeap =>
       raise Fail("Overread @{srcpos}")
      AllocatedHeader(ol,oc) |
      AllocatedPadding(ol,oc) => 
       raise Fail("Overread @{srcpos}: belongs to @{ol}")
      Allocated(ol,oc) => 
       if (oc != pc) then raise 
        Fail("Overread @{srcpos}: belongs to @{ol}")
      AllocatedDirty(ol,oc) => 
       if (oc != pc) then raise 
         Fail("Overread @{srcpos}: belongs to @{ol}")
       else logAndRecover(addr, "Check dumpster dive")
  NotHeapPointer => 
   if includesHeapLoc(lts) then
     raise Fail("Tampering @{srcpos}")
\end{verbatim}
\caption{Memory access rules in the HeapSafety Policy}
\label{fig:HeapSafetyPseudocode2}
\Description{Pseudocode of \tagrule{LoadT} rule for HeapSafety}
\end{figure}

\subsection{Detecting Potential Dumpster Diving}\label{sec:dumpsterdiving}

In cybersecurity, ``dumpster diving'' refers to recovering confidential or secret information discarded without proper data sanitation. This includes data left on a physical hard-copy \cite{1968-dumpster-fraud-Kalat21}, on discarded hardware like old laptops or hard-drives\cite{cyberdumpsterdiving}, or even in the heap by programs that are still executing. For performance reasons, most heap memory managers do not zero out memory dealloacted by \codevar{free()}. As a result, it is sometimes possible for a clever attacker to recover secrets, notably access tokens and secret authentication keys, left in the heap after the buffer containing them has been legally freed. According to the C standard, programmers should initialize memory obtained using \codevar{malloc()} before reading it (or use \codevar{calloc()}, which zeroes automatically). However, nothing prevents an attacker from simply allocating large chunks of memory through legal \codevar{malloc()} calls and going through the proverbial trash in memory. Programs rarely zero memory before \codevar{free()}. Even when they do, an optimizing compiler might remove user clean up code before \codevar{free()} because it regards it as ``dead code.'' Like their physical analogs, heap dumpster diving attacks are somewhat unpredictable; sometimes there are no secrets in the trash. However, dumpster diving is considered a valuable tactic to an active attacker seeking lateral movement in a network, such as from a foothold in a webserver to the valuable internal database server. 

Detecting this class of vulnerability requires more coordination between the fuzzer and the policy. We treat dumpster-dive
detection as part of HeapSafety, shown in Figures~\ref{fig:HeapSafetyPseudocode} and \ref{fig:HeapSafetyPseudocode2}.
\tagrule{LoadT}, \tagrule{StoreT}, and \tagrule{MallocT} are the relevant rules. The condition for the vulnerability is straightforward to express in a Tagged C policy: a read before the first write to allocated memory. \tagrule{MallocT} tags newly allocated memory bytes as \tagtype{AllocatedDirty} (instead of simply \tagtype{Allocated}), meaning it is legally assigned to the user, but still contains old data. 
\tagrule{StoreT} changes the tags on these bytes to \tagrule{Allocated} when they are first overwritten. If \tagtype{AllocatedDirty} is detected while reading memory (\tagrule{LoadT}) then a dumpster diving attack is possible.

While such an uninitialized heap read is always illegal (a bug), it is not always dangerous (a vulnerability). Users interested solely in bug detection could set the policy to simply failstop. Pipe-Cleaner allows more security-minded users---those only interested in vul\-ner\-a\-bil\-i\-ties---to refine the results through coordination between the fuzzer and the policy.  

It would be unacceptably slow for the policy to try and determine if the illegal behavior is a vulnerability at every read of every byte of memory. Instead, it defers that decision to the fuzzer. The \tagrule{StoreT} rule logs a message and includes the address. Since the Tagged C Interpreter is a software-only system, it can emit the values in memory to the fuzzer. (We note that this this would not be possible in the
Tagged C implementation based on PIPE hardware support that we envision for the future.)

Once the fuzzing run is finished, either via success, or failstop, the fuzzer determines if there were secrets in the uninitialized heap reads. In our proof-of-concecpt experimental implementation, the Pipe-Cleaner fuzzer looks for secret tokens using the regular expressions identified by Meli et al. \cite{meli2019badgit}. 

\subsection{Preventing Heap Address Leaks}\label{sec:heapaddrSIF}

Not all vulnerabilities in the heap involve illegal behavior; information leaks are perfectly legal from the perspective of the C standard and cause no damage by themselves. 
However, they can be leveraged in an exploit chain to dramatic effect. Address Space Layout Randomization (ASLR) is a popular mitigation for attacks on memory. 
It works by randomizing the layout of the major components of the address space (heap, stack, libc, globals, etc) on each run, so hard-coded addresses no longer work in attacks. ASLR can be defeated through disclosing (leaking) the address of a desired component, so that an exploit can proceed. 
Return-to-libc \cite{retlibc-Shacham04} attacks require a libc function address and analogous heap attacks require the attacker to have a heap address \cite{Rashidi-ASLR-exploitlist24}. Here we describe a Heap\-AddressSIF policy to detect heap address disclosures.

This type of legal but vulnerable behavior is not typically something fuzzers can detect. However, Secure Information Flow (SIF) techniques~\cite{Denning77:SecureInformationFlow} are ideally suited for detecting and characterizing this type of problem. Tagged systems are well suited to supporting SIF techniques while traditional sanitizers might struggle to do so. 

The key elements of our HeapAddressSIF policy are shown in Figure~\ref{fig:HeapAddressSIFPseudocode}.
To detect if addresses can be leaked, effectively bypassing ASLR’s abstraction, we use value tags to distinguish heap pointers
from all other values. The {\tt MallocT} rule sets the
value tag of the returned pointer ({\tt pt'}) to \tagtype{ProtectedPtr}; all other values are initialized to \tagtype{UnProtected}.
In our proof-of-concept implementation, the C interpreter 
only supports one way of putting out data, via \codevar{printf()}, which has its own control point and tag rule.
\tagrule{PrintfT} is passed a list of the value tags of the \codevar{printf()} arguments; if any of these is a heap pointer, the rule failstops.
(The scheme presented here is over-simplified in that it does not handle arguments formatted with {\tt \%s}, which \emph{should} be allowed
to be heap pointers, as this conversion specification causes the contents of the pointed-to string to be printed, rather than the address.)

\begin{figure}[tb]
  \centering
  \small
\begin{verbatim}
vtags := {UnProtected, ProtectedPtr}
  
Function MallocT() :=
 pt' :=  ProtectedPtr

Function BinopT(srcpos, vt1, vt2) :=
 case vt1, vt2 of
  ProtectedPtr, _ | _, ProtectedPtr =>
   vt' := ProtectedPtr
  UnProtected, UnProtected =>
   vt' := UnProtected

Function PrintfT(srcpos, arg_tags) :=
 for vt in arg_tags 
  if vt = ProtectedPtr then
   raise Fail("Address leak @{srcpos}")
\end{verbatim}
  \caption{Key Rules in the HeapAddressSIF Policy}
  \Description{Pseudocode of {\tt MallocT}, {\tt BinopT} and {\tt PrintfT} rules for HeapAddressSIF}
  \label{fig:HeapAddressSIFPseudocode}
\end{figure}

Similarly to the HeapSafety policy, the tag rules for arithmetic operations and casts propagate the \tagtype{ProtectedPtr} tag into
all result values. We show just the rule for binary operations (\tagrule{BinopT}) here; it consumes the
values tags on the two arguments ({\tt vt1},{\tt vt2}) and sets the value tag {\tt vt'} of the result to be \tagtype{ProtectedPtr} if
\emph{either} of the argument tags is. 
So, for example, converting a heap pointer value into its text representation (by performing shifts, masks, additions, etc.)
will ``taint'' the resulting character value tags and ultimately prevent them from being printed.

\section{Methodology and Metrics}
\label{sec:m-and-m}
\subsection{Metrics}\label{sec:metrics}

We measure our proof-of-concept's applicability to the three problems discussed in Section~\ref{sec:intro} as follows.

The Duplication problem is reasonably measured by the \emph{duplication rate}, the ratio of fuzzer-reported unique bugs to the number of unique ground-truth bugs actually in the report as determined by manual analysis. For example, if the fuzzer reports 3 bugs 
and manual triage determines there are actually only 2 (one was a duplicate), then the duplication rate would be 3/2. 
The ideal duplication ratio is 1, meaning no bugs were incorrectly duplicated by the fuzzer. If no bugs are detected, then the duplication ratio has no meaning. 

The Crash Bias problem does not lend itself to an easily quantified metric. Prior work seems to be more concerned with increasing overall numbers of bugs found rather than increasing the variety of bug classes found, although it is accepted that fuzzing benchmarks should have such variety~\cite{PrudentPracticeFuzzing24}. In targets containing multiple classes of bugs, we propose measuring fuzzer \emph{biodiversity} as the number of unique bug types detected. While characterizing bug classes, and relating specific vulnerabilities to classes of bugs, as is done by the MITRE CVE~\cite{cve-online} (vulnerabilities) and CWE~\cite{cwe-online} (weaknesses, or classes of bugs) is somewhat a matter of taste, we think the idea is sound for most targets. 

For example, suppose Fuzzer A reports finding 5 bugs, and manual triage determines that there is 1 heap overread (+1 duplicate), 3 different heap overwrites, and 1 unreproducible unknown. Fuzzer A’s true bug count is 4 and its biodiversity score is 2. Suppose fuzzer B reports 3 bugs, 1 heap overread, 1 heap overwrite, and 1 double free, with no duplicates.  Fuzzer B’s true bug count is 3, and its biodiversity score is 3. Fuzzer B performs better with respect to biodiversity, even though it found fewer overall bugs.  
For users interested in versatility and the Crash Bias problem, better diversity might be more desirable than a higher bug count. 
For targets containing only a single class of bugs, biodiversity is not an interesting metric because the maximum score is 1; these targets are at most demonstrations of novel detection abilities.

The Crash Triage problem is more difficult to quantify than the other two problems because bug triage is fundamentally very subjective. Therefore we prefer a qualitative approach to its assessment, such as performing user surveys that rate the usefulness of the output of Pipe-Cleaner vs. current fuzzers.

While target code coverage, i.e., how much of the target code runs, is a popular metric for fuzzers, it is orthogonal to our concerns. While exercising the code containing a bug is necessary for dynamic detection of the bug, it does not impact the three problems we consider.

\subsection{Experimental Configuration}
Our experiments compare the behavior of two fuzzers. The experimental fuzzer is Pipe-Cleaner, composed of the Tagged C custom policies, the Tagged C interpreter, the Interpreter’s VMF executor \& feedback modules, and VMF 3.1.0 \cite{VMF} augmented with the Pipe-Cleaner modules. The baseline fuzzer is composed of the Null Policy (which emulates a system without Tagged C policies), the Tagged C interpreter, the Interpreter’s VMF executor \& feedback modules, and base VMF 3.1.0. 

The fuzzing targets are specially crafted for this evaluation to work within the constraints of the Tagged C interpreter and contain known bugs. All of the existing benchmarks require features that the exploratory interpreter does not support, and do not exercise our novel bug detection capabilities (see also Section~\ref{sec:threatstovalidity}). The initial seed, or input, supplied to all fuzzers, is the uninformed seed, a single file containing the string ‘hello’.

Since we are interested in the detection, deduplication, and triage of bugs rather than test fuzzing coverage, and since this is a preliminary study on small targets, we limit our test runs to ten minutes rather than the recommended 24 hours \cite{Klees18:Evaluating, PrudentPracticeFuzzing24}. Although multiple cores are available, experiments are run individually to avoid RAM starvation. Experiments are repeated 30 times each, in keeping with best practice \cite{Klees18:Evaluating, PrudentPracticeFuzzing24}. Manual analysis is used to determine deduplication count and correctness of reported results. 

For completeness, we note the following details, though we think their influence is negligible due to limitations of the interpreter. There are eight Intel(R) Xeon(R) 3.50GHz CPUs, and 14 Gi of available RAM with 2 Gi of swap. The OS is Ubuntu 20.04. No Docker or virtualization is used.

\section{Results}
This is a Stage 1 submission and results are not included. Please see \url{https://dl.acm.org/journal/tosem/registered-papers} for details.

\section{Discussion}

The Pipe-Cleaner system demonstrates great potential for letting users easily customize their fuzzing runs to their interests and goals.

Since Pipe-Cleaner allows users the flexibility to design their own security policies, it will be interesting to see whether narrow policies or broad ones most benefit fuzzing and triage. Certain users, such as red teams, might prefer to focus narrowly on specific bug classes. In very focused policies, such as DoubleFree, which finds only two types of faults, messages can be tailored and triage becomes semi-automatic. Focused policies might require fewer and smaller error messages including less overhead. Other users might favor broader policies to detect more flaws. Broader policies might increase the overall yield of fuzzing, but the results might not be as easy to triage, as in HeapSafety, which finds at least five types of problems. Broader policies might also require bigger error messages, or incur a larger performance penalty.

\subsection{Preliminary Findings and Status}

The DoubleFree policy is fully built and integrated with the fuzzer, and gives good preliminary results. The basic fuzz targets show an ideal deduplication ratio of 1, and the experimental fuzzer identifies and discards duplicates robustly. Biodiversity is less meaningful since there are only two classes of violations in the basic fuzz targets, but Pipe-Cleaner does find them both for a score of 2.

The HeapSafety policy is fully finished, and its more involved fuzzer integration is in progress. We expect that for our highly constrained targets both the baseline fuzzer and the experimental one will ultimately find all the classic memory corruption vulnerabilities. We hope the experimental fuzzer has fewer duplicates and a reduced triage time. For the classic memory corruptions, we expect the diversity scores would be the same. We expect that for dumpster diving the experimental fuzzer will have reasonable deduplication and discard rates. We expect a higher diversity score than the baseline fuzzer for dumpster diving because the baseline fuzzer is highly unlikely to detect this class of attacks.

The HeapAddressSIF policy is in development. Once the policy is ready, its fuzzer integration is expected to be straightforward, nearly identical to DoubleFree. We expect that there will be reasonable deduplication and discard rates. We expect a higher diversity score than the baseline fuzzer because the baseline fuzzer is unlikely to detect this class of attacks.

\subsection{Implementation Limitations}\label{sec:implimits}
There are several major limitations to the current system that prevent support of realistic targets. C is not usually interpreted, because that is much slower than running compiled code. Popular binary coverage mechanisms in fuzzing have no meaning in the current interpreter. The lack of coverage mechanisms limits the fuzzer’s ability to make intelligent decisions during normal execution; it has to randomly keep a subset of inputs.  The interpreter’s stack and heap are much smaller than a realistic system. The interpreter can only handle one (small) source file and is single-threaded. The tag models for library behavior are limited to two functions, \codevar{ getchar()} and \codevar{printf()}. \codevar{fgets()} is present but has no tag support and any string functions must be implemented explicitly. There are no models for system calls. We expect that the interpreter will not be a long-term component of Pipe-Cleaner.
 
\subsection{Threats to Validity}\label{sec:threatstovalidity}
Because Pipe-Cleaner fuzzes for classes of bugs that have not been supported by other fuzzers, existing benchmarks cannot exercise its ability to find those bugs. Our proof-of-concept depends on a smaller set of hand-written examples. We can show that Pipe-Cleaner detects bugs in these examples, but they might not be representative of all the ways the bugs might appear in the wild. 

The other limitation of our prototype is its dependence on Tagged C, which is itself a young project. It might turn out that Tagged C cannot express policies that are worth fuzzing, or the cost of developing policies might be too high for non-experts. Tagged C cannot detect bugs such as integer overflow errors, and these might prove more important than the bugs it can find.

The system as a whole might not scale as expected. When fuzzing realistic, complex targets, speed does matter even if the results are improved. Whether the final performance tradeoff is worthwhile will depend on the user's goals. 

\section{Related Work}\label{relatedwork}
Pipe-Cleaner addresses the same goals as the broader fuzzing field, using concepts from property-based testing and dynamic security monitors. We briefly discuss the most relevant existing work, and then describe why we choose Tagged C and PIPE as our enforcement mechanism.  

\subsection{Fuzzing}
While no longer young, fuzzing remains an active field of research. The original fuzzer was somewhat naive \cite{OG-fuzz-Miller90}, but is still effective today \cite{30yrfuzzing-Miller2022}.  Open problems still remain, especially around the stability and consistency of the fuzzing results \cite{Fuzzingchalls-Boehme21,fuzzing-sota-Liang18,fuzzing-ASE-survey-Manes21}. The utility of results 

(or lack thereof) remains a top concern of fuzzing users \cite{Nourry-humanfuzzingchals23}. Prior work on duplication in fuzzing has focused on performing post-crash analysis, tracing, and clustering after normal fuzzing runs, rather than enriching fuzzing at the start as we do. AURORA\cite{AURORA-fuzz-Blazytko20} uses traces and delta debugging to group crashes,
IGOR\cite{IGOR-crashdedup-Juang21} uses traces and control flow graph similarity to group crashes, and FuzzerAid\cite{FuzzerAid-Joshy23} uses traces to generate code snippets, and heuristics to group crashes. 
Hardware tracing support has been shown to improve fuzzing performance~\cite{fuzzhardwaresupport-Ding21, PTrix-Chen19}. 

\subsection{Property-based Testing}
Property-based Testing (PBT) frameworks like QuickCheck \cite{Claessen00:QuichCheck} and QuickChick \cite{Paraskevopoulou15:QuickChick} resemble fuzzers in that they feed random inputs to a program, but instead of detecting crashes, they detect violations of hand-coded formal properties. They are generally used as a validation approach, lighter-weight than theorem proving; for instance, Lampropolous et al. \cite{Lampropoulos19:CoverageGuidedPBT} use PBT to rapidly validate a non-interference property and associated tag policy during ongoing development. PBT is also used to automatically generate test cases \cite{hypothesis-pbt}. PBT has yet to be fully applied to the use case of a typical fuzzer. The closest is PGFuzz, which takes a runtime-monitoring approach (see below) to fuzzing for violations of flight state invariants in drone flight control software \cite{kim2021pgfuzz}. PGFuzz's policies are expressed in temporal logic, and they describe potential problems in the system's external behavior, as opposed to describing risky internal behaviors of the code itself like our policies. Their system uses these policies to guide fuzzing, biasing their inputs to focus on bugs that have practical consequences.

\subsection{Runtime Monitoring}
Pipe-Cleaner’s design requires a general runtime security mechanism that can express a wide range of security concepts. Schneider \cite{Schneider00:Automata} models such mechanisms abstractly as {\em security automata}, separate machines that run in parallel with the primary computation. This concept is realized in a wide range of runtime verification approaches, with the policies themselves expressed as temporal logic formulae \cite{Chabot15:CTemporalAssertions} or regexes (i.e., state machines) on traces of manually defined “events” \cite{Ball02:slic, Havelund08:RVC}. Such systems then instrument their code with software monitors to enforce the policy. These would be viable alternatives to Tagged C in the Pipe-Cleaner model, but both temporal logic and regex languages are complex to write policies with and far removed from the host programming language, raising the barrier to entry.

At the same time, a number of hardware mechanisms have been proposed to assist in the runtime enforcement of security properties. Some, like CHERI \cite{Cheri-Watson15}, ARM PAC \cite{mitre-PAC-desc}, and Intel MPK \cite{MPK-survey-Park23} focus on the specific class of memory safety policies, making them too narrow for our purposes. Tag-based reference monitors like PIPE \cite{Dhawan14:PUMP} are more general. They can enforce a wide range of policies, including memory safety, compartmentalization, and forms of information flow control (IFC) \cite{micropolicies-tagmonitors-Amorim15,VerifiedSIF14}. Tag policies are often written at the assembly level, which is a usability issue, but recent work on Tagged C has enabled C source-level definition of policies \cite{TaggedC-RV23}. A Tagged C policy consists of instantiations of a fairly small number of “tag rules” that are closely connected to C language constructs, making policies easier to define.

Sanitizers are a popular way to augment fuzzer bug detection capabilities by making more conditions crash \cite{google-fuzz-beyond-membugs}, but they provide the fuzzer with no more information for deduplication than a standard crash does (though some sanitizers now leave a message for the user in the stack trace). 

Also, sanitizers cannot be run in tandem with each other \cite{sanitizers-Peko21},
whereas Pipe-Cleaner has the flexibility to run multiple policies simultaneously and independently of each other.

\section{Future Work}\label{sec:futurework}
Pipe-Cleaner as presented here is the first step of a journey. To help it reach its full potential we plan to add more interesting security policies, such as SQL injection or command injection detection by SIF, format string vulnerability detection, type confusion detection, and stack safety. In order to support more realistic fuzzing targets, the current Tagged C interpreter needs to be replaced with a more efficient
execution engine that can also model the tag behavior of calls to unknown library code.
Integrating support for coverage measurement is also desirable. Ultimately, we plan to compile Tagged C to
machine-level tagged code for the PIPE system, also including support for native execution.

\section{Conclusion}
Contemporary production fuzzing results are difficult to process effectively due to excessive noise from duplicated crashes and lack of information for effective triage. They are also biased towards classes of bugs that can easily manifest as crashes. We believe these problems stem from the same underlying cause: a profound lack of information about the conditions of a fault. We have introduced Pipe-Cleaner, which integrates developer-written security policies for a runtime reference monitor with fuzzing. The policy framework, Tagged C, is flexible and customizable, providing a detailed record of approximate root cause to the fuzzer and expressing security properties normally beyond the reach of current fuzzers. The proof-of-concept system appears to succeed on small targets, and is a promising approach to scale up in the future.

\begin{acks}

We’re grateful to Andrew Ruef, Arlen Cox, Adrian Herrera, Nathaniel Filardo, Roberto Blanco, Silviu Chiricescu, and Steve Vittitoe for their generous advice and technical insight. This work was supported by the National Science Foundation under Grant No. 2048499, Specifying
and Verifying Secure Compilation of C Code to Tagged Hardware. Allison Naaktgeboren is funded under the Draper Scholars Program.
\end{acks}

\bibliographystyle{ACM-Reference-Format}
\bibliography{references-base}

\end{document}